\documentstyle[preprint,tighten,aps,floats,amsfonts]{revtex}
\def\met{\mbox{${\hbox{$E$\kern-0.6em\lower-.1ex\hbox{/}}}_T$}} 
\def\D0{D\O}                            
\def\etal{{\sl et al.}}                 
\begin{document}
\preprint{\Large Fermilab--Pub--98/095--E}
\draft
\title{\Large\bf
A Search for Heavy Pointlike Dirac Monopoles}
\date{\large March 23, 1998}
%
%
\author{                                                                      
B.~Abbott,$^{31}$                                                             
M.~Abolins,$^{27}$                                                            
B.S.~Acharya,$^{46}$                                                          
I.~Adam,$^{12}$                                                               
D.L.~Adams,$^{40}$                                                            
M.~Adams,$^{17}$                                                              
S.~Ahn,$^{14}$                                                                
H.~Aihara,$^{23}$                                                             
G.A.~Alves,$^{10}$                                                            
N.~Amos,$^{26}$
% ...
\\                                                                            
\vskip 0.50cm                                                                 
\centerline{(D\O\ Collaboration)}                                             
\vskip 0.50cm                                                                 
}                                                                             
\address{                                                                     
\centerline{$^{1}$Universidad de los Andes, Bogot\'{a}, Colombia}             
\centerline{$^{2}$University of Arizona, Tucson, Arizona 85721}               
\centerline{$^{3}$Boston University, Boston, Massachusetts 02215}             
\centerline{$^{4}$Brookhaven National Laboratory, Upton, New York 11973}      
\centerline{$^{5}$Brown University, Providence, Rhode Island 02912}           
\centerline{$^{6}$Universidad de Buenos Aires, Buenos Aires, Argentina}       
\centerline{$^{7}$University of California, Davis, California 95616}          
\centerline{$^{8}$University of California, Irvine, California 92697}         
\centerline{$^{9}$University of California, Riverside, California 92521}      
\centerline{$^{10}$LAFEX, Centro Brasileiro de Pesquisas F{\'\i}sicas,        
                  Rio de Janeiro, Brazil}                                     
\centerline{$^{11}$CINVESTAV, Mexico City, Mexico}                            
\centerline{$^{12}$Columbia University, New York, New York 10027}             
\centerline{$^{13}$Delhi University, Delhi, India 110007}                     
\centerline{$^{14}$Fermi National Accelerator Laboratory, Batavia,            
                   Illinois 60510}                                            
\centerline{$^{15}$Florida State University, Tallahassee, Florida 32306}      
\centerline{$^{16}$University of Hawaii, Honolulu, Hawaii 96822}              
\centerline{$^{17}$University of Illinois at Chicago, Chicago,                
                   Illinois 60607}                                            
\centerline{$^{18}$Indiana University, Bloomington, Indiana 47405}            
\centerline{$^{19}$Iowa State University, Ames, Iowa 50011}                   
\centerline{$^{20}$Korea University, Seoul, Korea}                            
\centerline{$^{21}$Kyungsung University, Pusan, Korea}                        
\centerline{$^{22}$Institute of Nuclear Physics, Krak\'ow, Poland}            
\centerline{$^{23}$Lawrence Berkeley National Laboratory and University of    
                   California, Berkeley, California 94720}                    
\centerline{$^{24}$Louisiana Tech University, Ruston, Louisiana 71272}        
\centerline{$^{25}$University of Maryland, College Park, Maryland 20742}      
\centerline{$^{26}$University of Michigan, Ann Arbor, Michigan 48109}         
\centerline{$^{27}$Michigan State University, East Lansing, Michigan 48824}   
\centerline{$^{28}$Institute for Theoretical and Experimental Physics,        
                   Moscow, Russia}                                            
\centerline{$^{29}$Moscow State University, Moscow, Russia}                   
\centerline{$^{30}$University of Nebraska, Lincoln, Nebraska 68588}           
\centerline{$^{31}$New York University, New York, New York 10003}             
\centerline{$^{32}$Northeastern University, Boston, Massachusetts 02115}      
\centerline{$^{33}$Northern Illinois University, DeKalb, Illinois 60115}      
\centerline{$^{34}$Northwestern University, Evanston, Illinois 60208}         
\centerline{$^{35}$University of Notre Dame, Notre Dame, Indiana 46556}       
\centerline{$^{36}$University of Oklahoma, Norman, Oklahoma 73019}            
\centerline{$^{37}$University of Panjab, Chandigarh 16-00-14, India}          
\centerline{$^{38}$Institute for High Energy Physics, Protvino 142284,        
                   Russia}                                                    
\centerline{$^{39}$Purdue University, West Lafayette, Indiana 47907}          
\centerline{$^{40}$Rice University, Houston, Texas 77005}                     
\centerline{$^{41}$Universidade do Estado do Rio de Janeiro, Brazil}          
\centerline{$^{42}$University of Rochester, Rochester, New York 14627}        
\centerline{$^{43}$CEA, DAPNIA/Service de Physique des Particules,            
                   CE-SACLAY, Gif-sur-Yvette, France}                         
\centerline{$^{44}$Seoul National University, Seoul, Korea}                   
\centerline{$^{45}$State University of New York, Stony Brook,                 
                   New York 11794}                                            
\centerline{$^{46}$Tata Institute of Fundamental Research,                    
                   Colaba, Mumbai 400005, India}                              
\centerline{$^{47}$University of Texas, Arlington, Texas 76019}               
\centerline{$^{48}$Texas A\&M University, College Station, Texas 77843}       
}                                                                             

\maketitle
\newpage
\vspace*{2in}
\begin{abstract}
We have  searched for  central  production of a  pair of  photons with high
transverse  energies in $p\bar  p$ collisions  at $\sqrt{s} =  1.8$~TeV using
70~pb$^{-1}$  of  data  collected with  the D\O\  detector at  the Fermilab
Tevatron in  1994--1996. If they exist, virtual heavy pointlike Dirac monopoles
could rescatter pairs of nearly real photons into this final state via a box
diagram.  We observe no excess of events above background, and set lower 
95\% C.L. limits of 610, 870, or 1580 GeV/c$^2$ on the mass of a spin 0, 1/2, 
or 1 Dirac monopole.
\end{abstract}

\pacs{PACS numbers: 12.60.-i, 14.80.-j, 13.85.Rm\\[2cm]
{\large\it Submitted to Phys. Rev. Lett.}}
%

\newpage

One of the  open questions  of particle  physics is the  existence of Dirac
monopoles~\cite{Dirac,Schwinger},    hypothetical carriers  of the magnetic
charge proposed by P.~Dirac to symmetrize the Maxwell equations and explain
the quantization of  electric charge. If  such magnetic monopoles exist,
then the  elementary magnetic  and electric  charges ($g$  and $e$) must be
quantized according to the following formula:
\begin{equation}
    g = \frac{2\pi n}{e},\;\; n = \pm 1, \pm 2, ...,\label{eq:quant}
\end{equation}
where $n$ is an unknown  integer. Here we assume  that the elementary
electric charge is
that of an  electron. If  free  quarks exist,   Eq.~(\ref{eq:quant})
should be modified by replacing $e$ with $e/3$, which effectively increases
$g$ by a factor of three.

\begin{figure}[htb]
\vspace*{3.0in}                      
\includegraphics{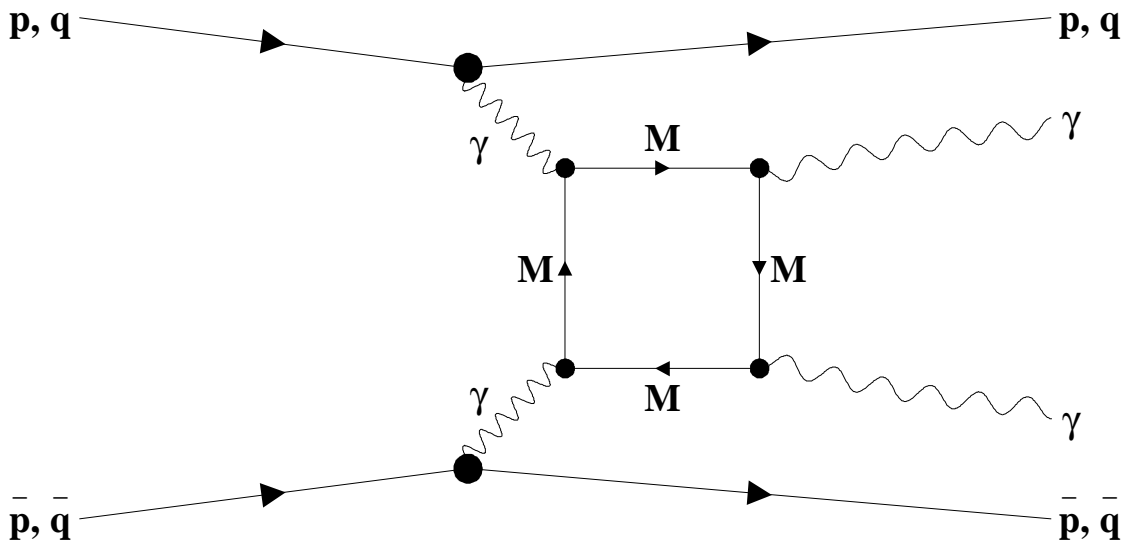}
\caption{
Feynman diagram for $\gamma\gamma$ production via a virtual monopole
loop.}
\label{fig:diagram}
\end{figure}

Dirac   monopoles are  expected  to  couple to  photons  with a
coupling constant $\alpha_g =  g^2/4\pi \approx 34\; n^2$ which is at least
three orders of magnitude larger  than the corresponding photon coupling to
the electric charge  ($\alpha_e = e^2/4\pi  \approx 1/137$). Therefore such
monopoles could give rise to photon-photon rescattering via the box diagram
shown in Fig.~\ref{fig:diagram}~\cite{Ginzburg1,DeRujula}. The contribution
of this diagram  for pointlike  monopoles to diphoton  production at hadron
colliders  was   recently     calculated~\cite{Ginzburg2}  and  shown to be
significant  even  for  monopole  masses  comparable to  the  collider beam
energy.

Since the  virtuality  ($Q^2$) of most  incoming photons  in the process of
Fig.~\ref{fig:diagram}  is   small~\cite{Ginzburg-private}, the interacting
partons  scatter at very  small  angles and  therefore escape  the detector
through the beam pipe. 
Thus, a  signature for monopoles at hadron colliders
is the   production of a  pair of  isolated  photons  with high  transverse
energies.  This process  gives a  unique  opportunity to find  evidence for
Dirac  monopoles  or to  set  limits on the  monopole  mass.

The only previous study of this nature was made by the L3 experiment at the
LEP $e^+e^-$  accelerator by  searching for the $Z \to  \gamma\gamma\gamma$
decay via a similar monopole  loop~\cite{L3}. It resulted in the lower 95\%
confidence  level  (C.L.)  mass  limit of  510~GeV for   pointlike spin 1/2
monopoles. Other accelerator  experiments (see Ref.~\cite{PDG}) have looked
for  production  of monopoles by  searching for the  high ionization traces
that  would  be  produced  by  these   particles, and  would   therefore be
inherently restricted to  monopole masses below  the beam energy. A variety
of experiments which look for monopoles in cosmic rays are sensitive to the
relic monopole  flux, rather  than the  monopole  mass~\cite{PDG}. Indirect
limits on the Dirac  monopole mass can be derived  from measurements of the
top quark mass  and the axial  and vector  couplings of the  $Z$ to charged
leptons~\cite{DeRujula}.

Despite  numerous  studies,  QED with  pointlike  monopoles is  still not a
complete theory. For example, it is  not clear whether such a theory can be
constructed  to be   renormalizable to  all   orders~\cite{DeRujula}. Also,
arguments  exist (see,   e.g.~\cite{Goldhaber})  that Dirac  monopoles must
occupy a spatial  volume of radius  $R \sim {\cal  O}(g^2/M)$, where $M$ is
the  monopole mass, to  accommodate  the self  energy implied  by the large
coupling. The non-observation of a  new distance scale in QED or the SM for
$R < {\cal O}$(1~TeV) requires the  monopole mass to exceed $\sim 100$~TeV.
Further  theoretical work on this  subject therefore is  required to define
the regions  of validity  for a  theory of  pointlike  monopoles. In such a
theory, it is  possible that hard  interactions of a  monopole with photons
would be weakened substantially by the effects of a monopole form factor.

In this Letter we report on the results of a new search for Dirac monopoles
with  the D\O\   detector   (described in  detail    elsewere~\cite{d0nim})
operating  at  the Fermilab  Tevatron  proton-antipoton  collider with beam
energies of  900~GeV. The  search is based  on $69.5 \pm  3.7$~pb$^{-1}$ of
data recorded in 1994--1996 using a  trigger which required the presence of
an electromagnetic (EM)  object with transverse  energy $E_T$ above 40~GeV.
This trigger did  not require the  presence of an  inelastic collision, and
therefore can be used to select low  $Q^2$ events typical of the process in
Fig.~\ref{fig:diagram}.

The following offline selection criteria are: (i) at least two photons with
$E_T >  40$~GeV and   pseudorapidity  $|\eta_\gamma|  < 1.1$;  (ii) missing
transverse  energy in  the event  $\met <  25$~GeV; and (iii)  no jets with
$E_T^j > 15$~GeV  and  $|\eta^j|<2.5$. The jet veto  requirement is used to
select the  low $Q^2$  process in   Fig.~\ref{fig:diagram}.  The trigger is
$>98$\% efficient for this off-line selection.

In order to  determine the  hard scattering  vertex, we  calculate the most
probable  direction of each  photon using  the transverse  and longitudinal
segmentation of the EM  calorimeter~\cite{Zvvg}. These directions determine
the position of the interaction  vertex along the beam axis. The resolution
on the  vertex position  for this  method is  14~cm, taken  from $Z \to ee$
decays where the  vertex can also be  determined with  high precision using
the tracking information. This  EM-cluster-based vertex finding techique is
preferred since for the  event topology of  Fig.~\ref{fig:diagram} one does
not expect charged particles,  causing the tracking-based vertex finding to
be biased  significantly toward  vertices from  background interactions. We
calculate  kinematic parameters of  the event based on  the vertex obtained
using the EM clusters.

Each  photon is  required to have:  (i) energy   isolation~\cite{Zvvg} $I <
0.1$;  (ii)  more than  95\% of  the  cluster  energy  deposited  in the EM
calorimeter;   (iii)  cluster shape   consistent with  that  expected for a
photon; and (iv) no  tracks pointing toward the  EM cluster from any of the
event vertices. 

The overall efficiency for photon  identification is $(73.0 \pm 1.2)\%$ per
photon, as  detailed in  Table~\ref{table:eff}. This  includes the ($92 \pm
1$)\% probability of the photon not  to convert in the material in front of
the tracking chambers. The  efficiency of criteria (i)--(iii) is determined
using the $Z  \to ee$ events  (with (ii)  additionally checked using a {\sc
geant}~\cite{GEANT}   simulation of the D\O\  detector for  possible energy
dependence); the efficiency of the no track requirement (iv) was determined
using simulated photons  obtained by rotating the  electromagnetic clusters
from $Z  \to ee$  decays by  $\pi/2$ in   azimuth~\cite{Zvvg}.  The overall
efficiency for the diphoton  selection is $(52.8 \pm 1.4)\%$. This includes
the   efficiency of  the  \met\ veto  $(99.0  \pm  0.5)$\%  as  well as the
identification efficiency for a pair of photons.

\begin{table}
\vspace*{0.15in}
\caption{Signal efficiency.}
\label{table:eff}
\begin{tabular}{ll}
Cut & Efficiency (\%) \\
\tableline
\multicolumn{2}{c}{per photon}\\
$I < 0.10$ & $93.0 \pm 0.7$\\
EM fraction & $99.0 \pm 1.0$\\
Shape consistency & $94.7 \pm 0.8$\\
No tracks & $91.1 \pm 0.4$\\
No photon conversions & $92.0 \pm 1.0$\\
\tableline
\multicolumn{2}{c}{per event}\\
$\met < 25$ GeV & $99.0 \pm 0.5$\\
\tableline
Overall & $52.8 \pm 1.4$ \\
\tableline
\end{tabular}
\end{table}
The  above  selection criteria  define  our base  sample which  contains 90
candidate events.

The main backgrounds to  photon scattering  through a monopole loop are due
to: (i) diagrams similar to  Fig.~\ref{fig:diagram} with other particles in
the  loop;  (ii)  QCD  production  of  dijets  ($jj$)  and  direct  photons
($j\gamma$) (with jets misidentified as photons due to fragmentation into a
leading $\pi^0$ or $\eta$ decaying  into a pair of spatially close photons,
reconstructed as one EM cluster), or direct diphotons ($\gamma\gamma$); and
(iii)  Drell-Yan  dielectron  production  with  electrons  misidentified as
photons due to tracking inefficiency.

Background (i) is dominated by a  virtual $W$-loop and has been shown to be
negligible~\cite{W-loop}.    The  other two  background   contributions are
estimated  from  the  data. The  QCD  background  is  determined  using the
$j\gamma$ event sample collected with a single photon trigger, with the jet
passing the same  fiducial and  kinematic cuts as the  photon. We apply a a
jet-faking probability $P(j \to  \gamma)$ which we measure to be $(10.5 \pm
1.5)\times 10^{-4}$ by  counting the number of  photons in multijet events,
and find  the QCD  background to be  $25 \pm 8$  events.  Direct photon and
diphoton  backgrounds are  also included in  this estimate.  Their relative
fractions are  obtained from {\sc   PYTHIA}~\cite{PYTHIA} Monte Carlo (MC).
The  30\% error   assigned to  the QCD   background  estimate  reflects the
uncertainty in  the direct  photon fractions  and in the  jet-faking-photon
probability. 

The Drell-Yan  background is calculated from a  sample of dielectron events
passing the same fiducial and kinematic cuts as the signal sample. Multijet
contamination of this sample is  negligible since the probability for a jet
to be  misidentified as an electron  is five times  smaller than that for a
photon. The  probability  for a  dielectron pair  to be  misidentified as a
diphoton  pair is found  to be $(11  \pm 1)\%$ by  comparing  the number of
events in the $Z$ peak in the $ee$ and $\gamma\gamma$ samples passing loose
kinematic cuts. The  Drell-Yan background in the  base sample is $63 \pm 7$
events. The  overall  background in the base  sample is $88  \pm 11$~(syst)
events, in good agreement with the 90 observed candidates. 

\begin{figure}[htb]
\vspace*{3.4in}                      
\includegraphics{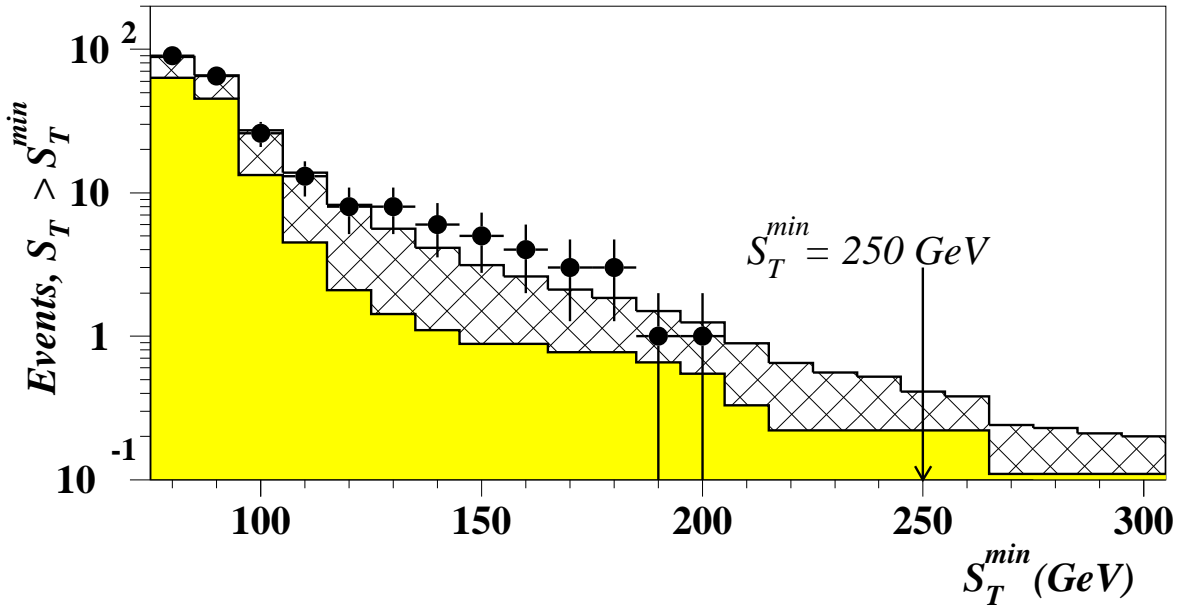}
\caption{
Data and  expected  background  as a  function of  $S_T^{\rm min}$ cut.
Points  are  data,  the upper  hatched  region   corresponds to  the  QCD
background, and the lower shaded region shows the Drell-Yan background. 
The $\approx 15$\% systematic error on the background is not shown.} 
\label{fig:bck}
\end{figure}

To   optimize  the   sensitivity  of  this  search  to  the   monopole loop
contribution we apply a cut on the scalar sum of the transverse energies of
all the photons in the event: $S_T \equiv \sum_i E_T^{\gamma_i}$~\cite{ST}.
We vary  the  $S_T$ cut  threshold   ($S_T^{\rm min}$)  in 10  GeV steps to
achieve   an  expected    background of  0.4     events~\cite{LQ}.  Such an
optimization is based on the fact that for this expected background one has
a 67\%  probability of  observing no  candidate  events in  the data in the
absence of a signal. In such a case~\cite{PDG}, the limits on the signal do
not  depend on  the  exact   background  value or  its   uncertainties. The
agreement   between  the   observed  number of   events and  the  predicted
background    as a   function  of    $S_T^{\rm  min}$  is    illustrated in
Fig.~\ref{fig:bck}.    Note that  since  the plot  shows  the  data and the
backgrounds for  $S_T > S_T^{\rm  min}$, the points are  highly correlated.
The $S_T^{\rm min} = 250$~GeV cut  corresponds to a background of $0.41 \pm
0.11$   events. The  event  in  the  base  sample  with  highest  $S_T$ has
$S_T=203$~GeV,   well  below this  cut. Taking  into account  the selection
efficiency we set an upper limit for the production cross section of two or
more photons with $\sum E_T^\gamma > 250$~GeV and $|\eta^\gamma| < 1.1$: 
\begin{equation}	
    \sigma(p\bar p \to \ge 2\gamma)|_{S_T >  
    250~\mbox{\tiny GeV},  |\eta^\gamma| < 1.1} < 83~\mbox{fb} 
\label{eq:limit}
\end{equation} 
at the 95\% C.L.  This limit is obtained using a Bayesian
approach with a flat prior and with the uncertainties in the efficiency and
the integrated luminosity properly taken into account.

\begin{figure}[htb]
\vspace*{3.3in}                      
\includegraphics{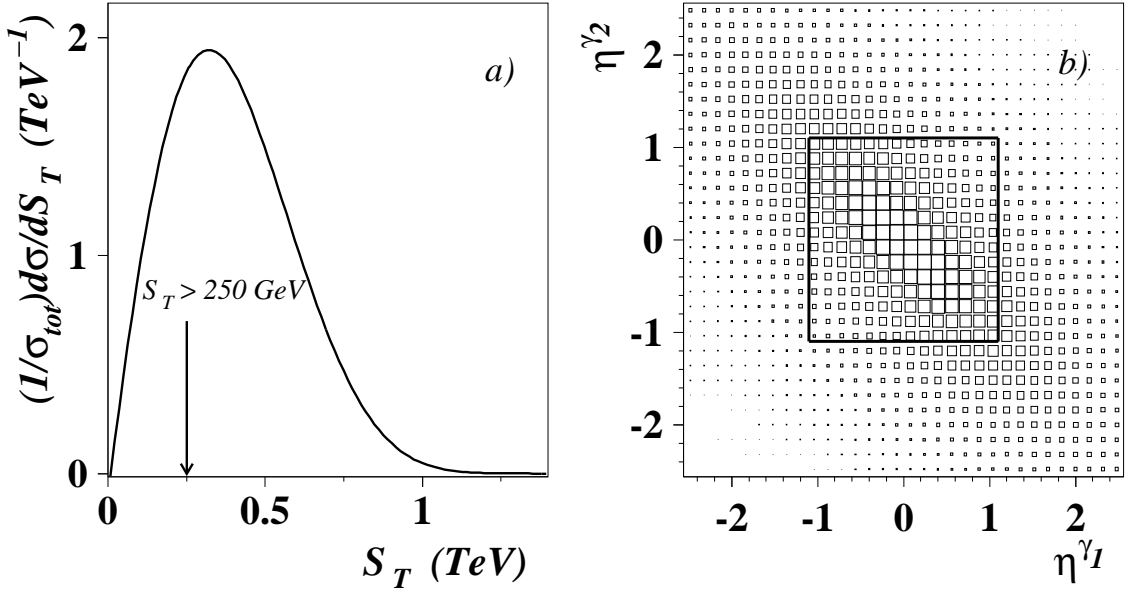}
\caption{a) Normalized $S_T$ spectrum and b) photon
pseudorapidities for the diphoton production via a heavy monopole loop.
The arrow in a) and square in b) show the chosen cuts in the corresponding
parameters.}
\label{fig:MC}
\end{figure}

Since the data  are consistent with  the background  hypothesis, we can set
limits on the  production of  pointlike  Dirac monopoles.  We calculate the
acceptance for the  monopole signal using a fast  MC program that generates
diphoton   events  from  a  monopole  loop   according  to  the  calculated
differential   cross  section     $d^3\sigma/dE_T^\gamma   d\eta^{\gamma_1}
d\eta^{\gamma_2}$~\cite{Ginzburg2}  with a subsequent parametric simulation
of the  D\O\  detector. The MC  model  takes into  account the  interaction
vertex distribution; parton density  distributions in the colliding protons
and  antiprotons, as  described by the  GRV~\cite{GRV}  parton distribution
functions  (p.d.f.); smearing of  photon momenta; and  detector acceptance.
Figures~\ref{fig:MC}(a) and (b) show the expected signal $S_T$ distribution
and the correlation between the  photon pseudorapidities, respectively. The
cuts  used in  this  analysis are  indicated  in the  figures.  The overall
acceptance for the monopole signal is found to be $(51 \pm 1)\%$, where the
error reflects variations due to  different p.d.f. (estimated by taking the
acceptance difference using GRV and  CTEQ4L~\cite{CTEQ4L}), and uncertainty
in the detector response parametrization. The acceptance does not depend on
the monopole mass  for masses above  the typical photon  energy ($\sim 300$
GeV)~\cite{Ginzburg-private}.

The total cross section of the heavy monopole production at the Tevatron is
given by~\cite{Ginzburg2}:
\begin{equation}
    \sigma(p\bar p \to \gamma\gamma + X) = 57\,P\; \left(\frac{n}{M~[{\rm
    TeV}]}\right)^8~{\rm fb,}
    \label{eq:cs}
\end{equation}
where $P$ is  a spin  dependent   factor~\cite{P-factor,W-loop}: P = 0.085,
1.39,  and 159  for  monopole  spin of  0, 1/2,  and 1,   respectively. The
estimated error on this cross section due to choice of p.d.f. and to higher
order QED effects is 30\%~\cite{Ginzburg-private}. Additional uncertainties
are  associated with  the  $\gamma\gamma  \to  \gamma\gamma$  subprocess in
Fig.~\ref{fig:diagram}   and with  unitarity  considerations.  The coupling
constant $\alpha_g$ is replaced with an effective coupling~\cite{Ginzburg2}
obtained by  multiplying  $\alpha_g$  by a factor   $(E^\gamma/M)^2$, where
$E^\gamma$ is the  photon energy,  typically 300~GeV at  the Tevatron. Both
unitarity  and  perturbation  theory  assumptions  are  satisfied when this
factor is $\ll 1$~\cite{Ginzburg1,Ginzburg2}.

\begin{figure}[htb]
\vspace*{6.5in}                      
\includegraphics{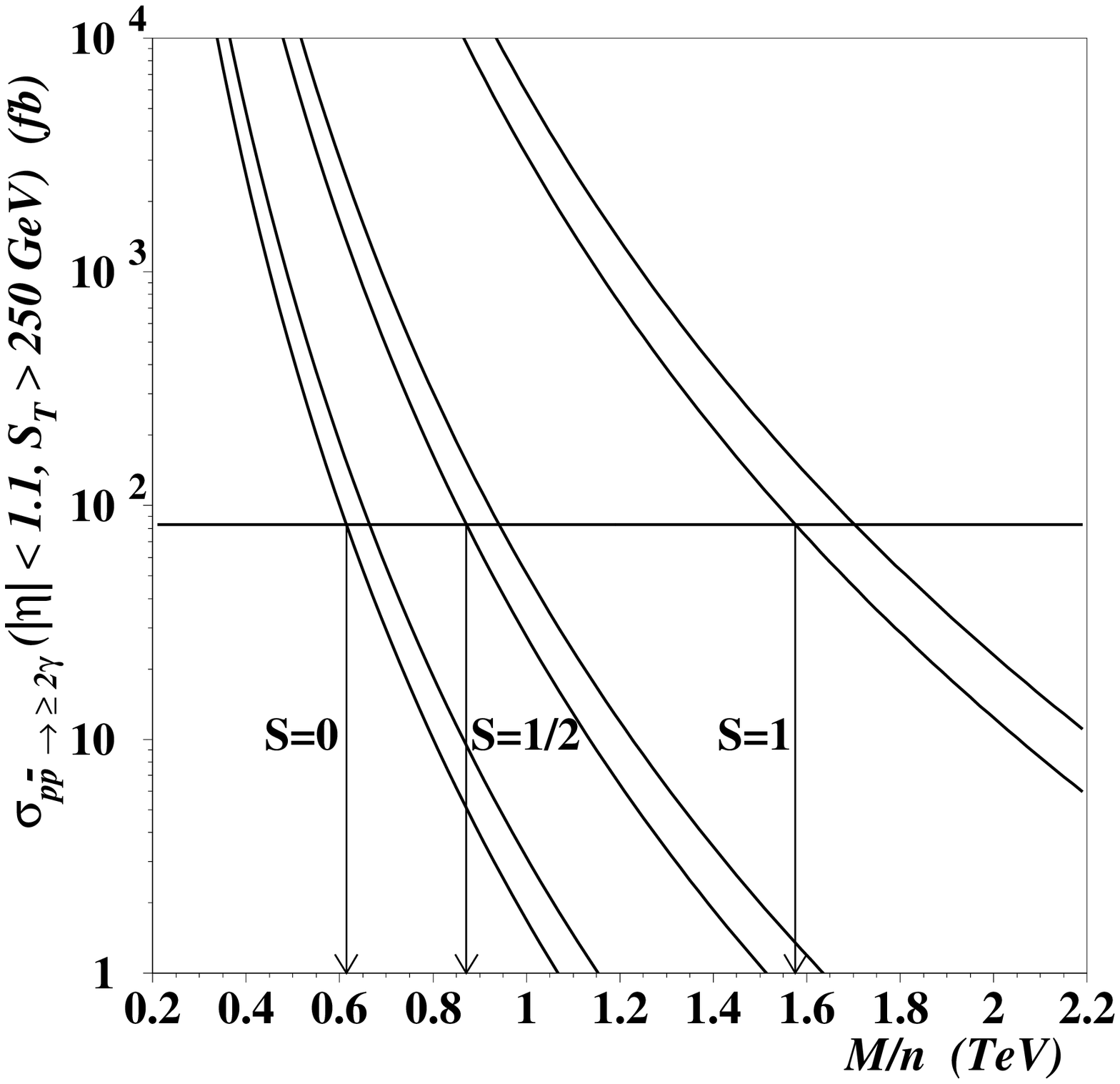}
\caption{The curved bands show
the low and upper bounds on theoretical cross sections~\protect\cite{Ginzburg2} 
for monopole spin, $S = 0, 1/2,$ and 1. The horizontal line shows the 95\% CL
experimental upper limit (\protect\ref{eq:limit}) on the cross section. The arrows indicate
the lower 95\% CL limits on the monopole mass at each spin value.}
\label{fig:limits}
\end{figure}

Comparing the  lower bound of the  theoretical cross  section corrected for
acceptance  with  the cross  section  limit   (\ref{eq:limit})  set by this
measurement, we  obtain the  following lower limits on  the pointlike Dirac
monopole mass (see Fig.~\ref{fig:limits}): 
$$
M/n > \left\{
    \begin{array}{rl}
     610~\mbox{GeV} & \mbox{for~} S = 0 \\
     870~\mbox{GeV} & \mbox{for~} S =  1/2  \\ 
    1580~\mbox{GeV} & \mbox{for~} S =  1
    \end{array}\right..  
$$
These are currently the most stringent limits on heavy pointlike monopole mass.
(We do assume, if more  than one type of Dirac  monopole exists, that there
is no cancellation among the loop diagrams involving each monopole type.)

We note that the effective coupling exceeds 1 and unitarity is violated
close to the  experimental  bound. For  values $E^\gamma/M  > 1$, the cross
section will grow more slowly,  approaching the usual $1/M^2$ behavior of a
QED  process~\cite{Ginzburg-private}  which satisfies  unitarity. Also, for
lower monopole  masses the effective  parameter of the  perturbation theory
used in the  calculations~\cite{Ginzburg2} becomes too large, and therefore
one would expect a non-negligible contribution of the higher order diagrams
with four,  six, etc.  photons in  the final  state. The  latter effect is,
however, largely  compensated by our analysis cut  on the sum of the photon
transverse  energies; if  part of  the signal  cross section  is due to the
higher order diagrams, the above limits are unaffected. 

When more complete  theoretical calculations are  available,  limits on the
monopole  mass could be  updated by  comparing  the modified  cross section
expression  with the   experimental limit   (\ref{eq:limit}).  With current
theory~\cite{Ginzburg2}    the above  limits  are  strictly  valid only for
monopole masses above several hundred GeV.

As a cross-check of the results of this search we have selected  elastic or
nearly elastic collisions by requiring no hits in the forward scintillating
hodoscopes used for  luminosity monitoring and  triggering on the inelastic
collisions~\cite{d0nim}.     This    requirement   drastically  reduces the
backgrounds. The remaining background in the base sample for elastic events
is $1.8 \pm  0.4$ events,  dominated by  diffractive  Drell-Yan events  and
residual    inelastic   background  due to    inefficiency of  the  forward
hodoscopes. We observe  one candidate event in  the base sample, consistent
with this expected background rate. For $S_T^{\rm min} \approx 100$~GeV the
background  is 0.4  events, and  no  candidates are  observed.  We use this
sample only  as a cross  check  because the  efficiency of  these selection
requirements is significantly lower  than that of the main analysis method,
primarily because of multiple interactions.

In   conclusion, we  have  performed a  search  for heavy   pointlike Dirac
monopoles by searching for pairs of  photons with high transverse energies.
Our  data  agree with  the  expected   background  from QCD  and  Drell-Yan
production.   No   candidates  pass  the  final  cuts.  Using   theoretical
calculations~\cite{Ginzburg2}  we  set 95\% C.L. lower  limits on the Dirac
monopole  mass for  minimum  magnetic  charge  $(n=1)$ in the  range 610 to
1580~GeV, depending on the monopole spin. These are the most stringent mass
limits on heavy pointlike Dirac  monopoles to date. Our cross section limit
(\ref{eq:limit})  is 83~fb, and may  be applicable to  the other production
processes, such  as that of  dyons~\cite{DeRujula} or  other exotic objects
strongly coupled to photons.

We are  grateful to  I.~Ginzburg and  A.~Schiller for many  discussions and
detailed  cross  section   information  and to  U.~Baur,   B.~Dobrescu, and
A.S.~Goldhaber for helpful discussions.
We thank the staffs at Fermilab and collaborating institutions for their
contributions to this work, and acknowledge support from the 
Department of Energy and National Science Foundation (U.S.A.),  
Commissariat  \` a L'Energie Atomique (France), 
State Committee for Science and Technology and Ministry for Atomic 
   Energy (Russia),
CAPES and CNPq (Brazil),
Departments of Atomic Energy and Science and Education (India),
Colciencias (Colombia),
CONACyT (Mexico),
Ministry of Education and KOSEF (Korea),
and CONICET and UBACyT (Argentina).


\begin{references}
%
\bibitem[*]{ecuador}
Visitor from Universidad San Francisco de Quito, Quito, Ecuador.

\bibitem[\dag]{beijing}
Visitor from IHEP, Beijing, China.

\vskip 0.25cm
%
\bibitem{Dirac}
P.A.M.~Dirac, Proc. R. Soc. London, Ser. {\bf A} 133, 60 (1931).
\bibitem{Schwinger}
J.~Schwinger, Phys. Rev. {\bf 151}, 1055 (1966).
\bibitem{Ginzburg1}
I.F.~Ginzburg, S.L.~Panfil, Sov. J. Nucl. Phys. {\bf 36}, 850 (1982).
\bibitem{DeRujula}
A.~De R\'ujula, Nucl. Phys. {\bf B435} 257 (1995).
\bibitem{Ginzburg2}
I.F.~Ginzburg and A.Schiller, {\tt hep-ph/9802310}, submitted to Phys. Rev.
D.
\bibitem{Ginzburg-private}
I.F.~Ginzburg, private communication.
\bibitem{L3}
L3 Collaboration, M.~Acciarri \etal, Phys. Lett. {\bf B345}, 609 (1995).
\bibitem{PDG}
PDG Review of Particle Physics, Phys. Rev. D {\bf 54}, 166, 685-687 (1996).
\bibitem{Goldhaber}
A.S.~Goldhaber, in Proceedings of the CRM--FIELDS--CAP
Workshop ``Solitons'' at Queen's University, Kingston, Ontario, July 1997
(Springer, New York 1998).
\bibitem{d0nim}
D\O\  Collaboration, S.~Abachi  \etal, Nucl. Instrum.  Methods {\bf A338},
185 (1994).
\bibitem{Zvvg}
D\O\  Collaboration, S.~Abachi  \etal, Phys. Rev. Lett. {\bf 78}, 3640
(1997); {\it ibid.} Phys. Rev. D {\bf 56} 6742 (1997).
\bibitem{GEANT} 
R.~Brun  and   F.~Carminati,  CERN  Program   Library  Writeup  W5013, 1993
(unpublished). We used {\footnotesize{GEANT}} version 3.15.
\bibitem{W-loop}
G.~Jikia and A.~Tkabaladze,    Phys. Lett.  {\bf  B323},  453  (1994).
\bibitem{PYTHIA}
T.~Sj\"ostrand,   Comp.    Phys. Comm.  {\bf  82},  74   (1994).  We used
{\footnotesize PYTHIA} version 5.7.
\bibitem{ST}
We use the $S_T$ variable instead of the individual photon energies,
since it is less sensitive to the next-to-leading order QED corrections to
the diagram in Fig.~\protect\ref{fig:diagram}. See discussion later in the
main text.
\bibitem{LQ}
D\O\  Collaboration, S.~Abachi  \etal, Phys. Rev. Lett. {\bf 79}, 4321 (1997);
B.~Abbott \etal, Phys. Rev. Lett. {\bf 80}, 2051 (1998).
\bibitem{GRV}
M.~Gl\"uck, E.~Reya and A.~Vogt, Z. Phys. {\bf C67}, 433 (1995).
\bibitem{CTEQ4L} 
CTEQ Collaboration, H.L.~Lai \etal, Phys. Rev. D {\bf 55}, 1280 (1997).
\bibitem{P-factor}
W.~Heisenberg and H.~Euler, Z. Phys.  {\bf 38}, 714 (1936); V.~Constantini,
B. De Tollis, and G.~Pistoni, Nuovo Cim. {\bf 2A}, 733 (1971);   M.~Baillagreon,
G.~Belanger,   and    F.~Boudjema,  Phys.  Rev. D  {\bf  51},  4712  (1995);
M.~Baillagreon,  F.~Boudjema, E.~Chopin, and  V.~Lafage, Z. Phys. {\bf C67},
431 (1996).
\end{references}
\end{document}